\documentclass[12pt]{article}

\usepackage[left=1.5in, right=1in, top=1in, bottom=1in, dvips,pdftex]{geometry}
\usepackage{color, setspace, graphicx, verbatim, mathrsfs, amsmath, amsthm, amssymb, amsfonts,  multirow, moreverb}
\newtheorem{defn}{Definition}[section]

\newtheorem{lemma}[defn]{Lemma}
\newtheorem{thm}[defn]{Theorem}

\theoremstyle{remark}

\numberwithin{equation}{section}
 \numberwithin{figure}{subsection}

\DeclareMathOperator{\airy}{Ai}

\def\ra{\rightarrow}
\def\iy{\infty}

\def\be{\begin{equation}}
\def\ee{\end{equation}}
\def\ov{\over}

\newcommand{\bP}{\mathbb{P}}

\begin{document}

\title{\textbf{Edgeworth Expansion of the Largest Eigenvalue Distribution Function
 of GUE Revisited }}

\author{Leonard N.~Choup \\ Department of Mathematics \\ University Alabama in Huntsville\\
 Huntsville, AL 35899, USA \\
 email:  \texttt{Leonard.Choup@uah.edu}
} \maketitle

\begin{center} \textbf{Abstract} \end{center}
\begin{small}
We derive expansions of the resolvent $R_{n}(x,y;t)=
(Q_{n}(x;t)P_{n}(y;t)-Q_{n}(y;t)P_{n}(x;t))/(x-y)$ of the Hermite
kernel $K_{n}$ at the edge of the spectrum of the finite $n$
Gaussian Unitary Ensemble (GUE$_n$) and the finite $n$ expansion of
$Q_{n}(x;t)$ and $P_{n}(x;t)$.
Using these large $n$ expansions, we give another proof of the
derivation of an Edgeworth type theorem for the largest eigenvalue
distribution function of GUE$_n$. We conclude with a brief
discussion on the derivation of the probability distribution
function of the corresponding largest eigenvalue in the Gaussian
Orthogonal Ensemble (GOE$_n$) and Gaussian Symplectic Ensembles
(GSE$_n$).
\end{small}
\section{Introduction}\label{intro}
The author stressed in \cite{Choup1} the importance of having a
large $n$-expansion of the distribution of the largest eigenvalue
from classical Random Matrix Ensembles. In this paper we present
another derivation of the probability distribution function of the
largest eigenvalue from the GUE$_{n}$. Unlike the previous
derivation which follows from the Fredholm determinant
representation $\bP(\lambda_{Max}\leq t)= \det(I-K_{n})_{(t,\iy)}$,
this one follows directly from the resolvent kernel representation
$\bP(\lambda_{Max}\leq t)= \exp \{
-\int_{t}^{\iy}R_{n}(x,x;t)\,dx\}$.
(A proof of this representation can be found in \cite{Trac7}.) Here
the Fredholm determinant expansion is replaced by the large
$n$-expansion of $P_{n}$ and $Q_{n}$. In doing this we discover new
integrals relating  Painlev$\acute{e}$ functions appearing in the
study of the largest eigenvalue in Gaussian Ensembles. These large
$n$-expansions can be used for the analogous problem of finding the
probability distribution of the largest eigenvalue in the GOE$_{n}$
and the GSE$_{n}$ case.

Recall that for Gaussian Ensembles, the probability density that the
eigenvalues are in infinitesimal intervals about the points $x_{1},
\cdots , x_{n}$ is given by
\begin{equation}\label{jpdfeig}
\bP_{n\beta}(x_{1},\cdots,x_{n})\; = \textrm{C}_{n\beta}\,
\textrm{exp}\left(-\frac{\beta}{2}\, \sum_{1}^{n}x_{j}^{2}\right)\,
\prod_{j<k}|x_{j}-x_{k}|^{\beta},
\end{equation}
with
\begin{equation}
-\infty < \lambda_{i} < \infty, \quad \textrm{for} \; i=1, \cdots,
n,
\end{equation}
and $\textrm{C}_{n\beta}$ is the normalization constant.\\
Let
\begin{equation}\label{p.d.f}
F _{n,\beta}(t)= \bP(\lambda_{\textrm{max}}^{\beta} \leq t)
\end{equation}
be the probability distribution function of the largest eigenvalue
in
 GOE$_n$ for $\beta\, =\, 1$,   GUE$_n$ for $\beta\, =\, 2$,  and
 GSE$_{n}$ for $\beta\, =\, 4$ respectively.\\
For the Gaussian $n$ Ensemble, the expected value of the largest
eigenvalue is asymptotically $\sqrt{2n}$. Therefore as the size of
the matrices grows, so does the largest eigenvalue. To have a
nontrivial limit, we must center and normalize
 $\lambda_{\textrm{max}}^{\beta}$. In doing this we keep the fine tuning constant
 $c$ introduced in \cite{Choup1},
  \begin{equation}
 \hat{\lambda}_{\textrm{max}}^\beta := { \lambda_{\textrm{max}}^G-\left(2(n+c)\right)^{1/2}\ov
 2^{-1/2} n^{-1/6}}.
 \end{equation}
 To state our results we need the following definitions.
Recall that if
\begin{equation*}
 \varphi_{n}(x)= {1\over (2^{n}n! \sqrt{\pi})^{1/2}} \, H_n(x)\,  e^{-x^2/2}
\end{equation*}
with $H_n(x)$  the Hermite polynomials of degree $n$, then the
Hermite kernel is
\begin{equation*}
K_{n}(x,y)=\sum_{k=0}^{n-1}\varphi_{k}(x)\varphi_{k}(y).
\end{equation*}
The resolvent of the integral operator on $L^{2}(t,\infty)$ with
 Hermite kernel will be denoted by $R_{n}$ and its kernel
denoted by
\begin{equation}
R_{n}(x,y):= (I-K_{n})^{-1}\,K_{n}\,(x,y).
\end{equation}
This resolvent also has the representation (see for example
\cite{Trac7}, page 6)
\begin{equation}\label{Rn}
R_{n}(x,y;t)\, =\,
\frac{Q_{n}(x;t)\,P_{n}(x;t)\,-\,P_{n}(x;t)\,Q_{n}(y;t)}{x-y}
\end{equation}
where
\begin{equation}\label{Qn}
Q_{n,i}(x;t)\, =\, (\,(I-K_{n})^{-1}\, ,\, x^{i}\varphi_{n})
\end{equation}
and
\begin{equation}\label{Pn}
P_{n,i}(x;t)\, =\, (\,(I-K_{n})^{-1}\, ,\, x^{i}\varphi_{n-1}).
\end{equation}
We introduce the following quantities
\begin{equation}\label{qn}
q_{n,i}(t)\>=\> Q_{n,i}(t;t), \>\>\> p_{n,i}(t)\>=\> P_{n,i}(t;t)
\end{equation}
\begin{equation}\label{un}
u_{n,i}(t)\,=\, (Q_{n,i},\varphi_{n}),\quad v_{n,i}(t)\,=\,
(P_{n,i},\varphi_{n}),\quad
\end{equation}
\begin{equation}
\tilde{v}_{n,i}(t)=(Q_{n,i},\varphi_{n-1}),\quad \mathrm{and} \quad
w_{n,i}(t)\,= \, (P_{n,i},\varphi_{n-1}).
\end{equation}
 Here $(\,\cdot \,,\cdot \,)$ denotes
the inner product on $L^{2}(t,\infty)$. In our notation, the
subscript without the $n$ represents the scaled limit of that
quantity when $n$ goes to infinity, and  we dropped the second
subscript $i$ when it is zero.\\  If $\airy$ is Airy function, the
kernel $K_{n}(x,y)$ then scales\footnote{For the precise definition
of this scaling, see the next section} to the  Airy kernel
\begin{equation}\label{notation}
K_{\airy}(X,Y)\>=\>
\frac{\airy(X)\,\airy^{'}(Y)\>-\>\airy(Y)\,\airy^{'}(X)}{X-Y}.
\end{equation}
Our convention is as follow;
\begin{equation}\label{Q}
Q_{i}(x;s)\, =\, (\,(I-K_{\airy})^{-1}\, ,\, x^{i}\airy),
\end{equation}
\begin{equation}\label{P}
P_{i}(x;s)\, =\, (\,(I-K_{\airy})^{-1}\, ,\, x^{i}\airy^{'}),
\end{equation}
\begin{equation}\label{q}
q_{i}(s)\>=\> Q_{i}(s;s), \>\>\> p_{i}(s)\>=\> P_{i}(s;s),
\end{equation}
\begin{equation}\label{u}
u_{i}(s)\,=\, (Q_{i},\airy),\quad v_{i}(s)\,=\, (P_{i},\airy),
\end{equation}
\begin{equation}
\tilde{v}_{i}(s)=(Q_{i},\airy^{'}),\quad \mathrm{and} \quad
w_{i}(t)\,= \, (P_{i},\airy^{'}).
\end{equation}
 Here $(\,\cdot \,,\cdot \,)$  denotes
the inner product on $L^{2}(s,\infty)$  and $i=0,1,2.$ \\
We use the subscript $n$ for unscaled quantities only.

Our first result are large $n$-expansions of $R_{n}(x,y;t)$,
$Q_{n}(x;t)= Q_{n,0}(x;t)$ and $P_{n}(x;t)=P_{n,0}(x;t)$.

\begin{thm}\label{resolvent kernel}
For
\begin{equation}\label{scaling}
x=\sqrt{2(n+c)}+ \frac{X}{2^{\frac{1}{2}}n^{\frac{1}{6}}}, \>\>
y=\sqrt{2(n+c)}+\frac{Y}{2^{\frac{1}{2}}n^{\frac{1}{6}}}\>\>
\mathrm{and} \>\> t
=\sqrt{2(n+c)}+\frac{s}{2^{\frac{1}{2}}n^{\frac{1}{6}}},
\end{equation}
as $n\rightarrow \iy$ with $X$, $Y$, and $s$ bounded,
\begin{equation*}
R_{n}(x,y;t)dx= \left[R(X,Y;s)-c\,Q(X;s) Q(Y;s)\,n^{-\frac{1}{3}}\,
+ \frac{n^{-\frac{2}{3}}}{20} \biggl[P_{1}(X;s) P(Y;s) +P(X;s)
P_{1}(Y;s)\right.
\end{equation*}
\begin{equation*}
- Q_{2}(X;s) Q(Y;s) - Q_{1}(X;s) Q_{1}(Y;s) -
 Q(X;s)  Q_{2}(Y;s) + 20 c^{2} u_{0}(s) Q(X;s) Q(Y;s)
\end{equation*}
\begin{equation}
  +\left. \left. \frac{3-20c^{2}}{2}\left(P(X;s)  Q(Y;s) + Q(X;s)
P(Y;s)\,\right) \right] + O(n^{-1})e_{n}(X,Y)\right]dX.
\end{equation}
The error term, $ e_{n}(X,Y)$, is the kernel of an integral operator
 on $L^{2}(s,\infty)$ which is trace class.
\end{thm}
And
\begin{thm}\label{Q_{n}}
For
\begin{equation*}
x=\sqrt{2(n+c)}+ \frac{X}{2^{\frac{1}{2}}n^{\frac{1}{6}}} \quad
\mathrm{and} \quad t=\sqrt{2(n+c)}+
\frac{s}{2^{\frac{1}{2}}n^{\frac{1}{6}}},
\end{equation*}
as $n\rightarrow \iy$ with $X$ and $s$ bounded,
\begin{equation*}
Q_{n}(x;t)=n^{\frac{1}{6}}\biggl[ Q(X;s)+
\left[\frac{2c-1}{2}P(X;s)-c Q(X;s)u(s)\right]n^{-\frac{1}{3}}
\end{equation*}
\begin{equation*}
+\left[(10c^{2}-10c+\frac{3}{2})Q_{1}(X;s)+P_{2}(X;s) +
(-30c^{2}+10c+\frac{3}{2})Q(X;s) v(s) \right.
\end{equation*}
\begin{equation*}
+  P_{1}(X;s) v(s) +P(X;s) v_{1}(s)-Q_{2}(X;s) u(s)-Q_{1}(X;s)
u_{1}(s)-Q(X;s) u_{2}(s)
\end{equation*}
\begin{equation}
+ \left.(-10c^{2}+\frac{3}{2})P(X;s) u(s) +20c^{2}Q(X;s) u^{2}(s)
\right]\frac{n^{-\frac{2}{3}}}{20} +O(n^{-1})E_{q}(X;s)\biggr],
\end{equation}
and
\begin{equation*}
P_{n}(x;t)=n^{\frac{1}{6}}\biggl[ Q(X;s)+
\left[\frac{2c+1}{2}P(X;s)-c Q(X;s)u(s)\right]n^{-\frac{1}{3}}
\end{equation*}
\begin{equation*}
+\left[(10c^{2}+10c+\frac{3}{2})Q_{1}(X;s)+P_{2}(X;s) +
(-30c^{2}-10c+\frac{3}{2})Q(X;s) v(s) \right.
\end{equation*}
\begin{equation*}
+  P_{1}(X;s) v(s) +P(X;s) v_{1}(s)-Q_{2}(X;s) u(s)-Q_{1}(X;s)
u_{1}(s)-Q(X;s) u_{2}(s)
\end{equation*}
\begin{equation}
+ \left.(-10c^{2}+\frac{3}{2})P(X;s) u(s) +20c^{2}Q(X;s) u^{2}(s)
\right]\frac{n^{-\frac{2}{3}}}{20} +O(n^{-1})E_{p}(X;s)\biggr].
\end{equation}
\end{thm}

Theorem \ref{Q_{n}} together with the work of Tracy and Widom in
\cite{Trac7}, all gives another proof of the following large
$n$-expansion of \eqref{p.d.f} when $\beta\,=\,2$.
\begin{thm}\label{GUE}
We set
\begin{equation}\label{scale for GUE}
t  =  (2(n+c)
)^{\frac{1}{2}}+2^{-\frac{1}{2}}n^{-\frac{1}{6}}\,s\>\>\>
\textrm{and}
\end{equation}
\begin{equation}\label{second term}
E_{c,2}(s)=2w_{1}-3u_{2}+ (-20c^2 +3)v_{0}  + u_{1}v_{0}-u_{0}v_{1}
+u_{0}v_{0}^{2}-u_{0}^{2}w_{0}.
\end{equation}
Then as $n\ra\iy$
\begin{equation}\label{GUE Edgeworth}
F_{n,2}(t)= F_{2}(s)\left\{ 1 + c \, u_{0}(s) \, n^{-\frac{1}{3}}
-\frac{1}{20}E_{c,2}(s)\, n^{-\frac{2}{3}}\right\} + O(n^{-1})
\end{equation}
uniformly in $s$, and
\begin{equation}\label{TW}
F_{2}(s)\>=\> \lim_{n\ra\iy}F_{n,2}(t)\>=\>
\exp\left(-\int_{s}^{\iy}(x-s)q(x)^2 \,dx\right)
\end{equation}
is the Tracy-Widom distribution.
\end{thm}

In \S 2 we derive  Theorem \ref{resolvent kernel} and Theorem
\ref{Q_{n}}. In \S 3 we give another proof of \eqref{GUE Edgeworth}
where this time we make use of the representation of the probability
distribution function of the largest eigenvalue in term of the
resolvent of the Hermite kernel instead of the Fredholm determinant
representation using the Hermite kernel. The advantage of this
derivation is that this technique also applies to the finite $n$
GOE$_{n}$ and GSE$_{n}$.

\section{Large $n$-Expansion of $R_{n}(x,y;t),\>\> Q_{n}(x;t)$ and
$P_{n}(x;t)$}

In this section we will make use of the following expansion of the
Hermite kernel $K_{n,2}(x,y)$  derived in \cite{Choup1}. Let
$\airy(x)$ be the Airy function and
\begin{equation}
K_{\airy }(x,y)\>= \>\frac{\airy(x) \airy^{'}(y) - \airy(y)
\airy^{'}(x)}{x-y}=\int_0^{\iy} \airy(x+z)\airy(y+z)\, dz
\label{airyKernel}
\end{equation}
the Airy kernel. For $x$ and $y$  defined by \eqref{scaling}, we
have as $n\ra\iy$
\begin{equation*}
K_{n,2}(x,y)\,dx =  \left\{  K_{\airy}(X,Y) -c_{G}\airy(X)\airy(Y)
n^{-\frac{1}{3}} +   \right.
\end{equation*}
\begin{equation*}
\frac{1}{20}\left[(X + Y)\airy^{\prime}(X)\airy ^{\prime}(Y) -
(X^2+XY+Y^2)\airy(X)\airy(Y) + \right.
\end{equation*}
\begin{equation}\label{hermite kernel}
\left. \left. \frac{-20c_{G}^2 +3 }{2}\left(\airy
^{\prime}(X)\airy(Y) + \airy(X)\airy ^{\prime}(Y)\right)
\right]n^{-\frac{2}{3}}  +O(n^{-1}) E(X,Y) \right\} \, dX,
\end{equation} uniformly in $s$.
The error term, $ E(X,Y)$, is the kernel of a trace class integral
operator on $L^{2}(s,\iy)$.

In order to simplify the notations in this paper, we treat each term
appearing in \eqref{hermite kernel} as an integral operator as well
as the kernel of that operator. For example $\airy(x)\airy(y)$ will
be the integral operator with this kernel.

We recall that the resolvent operator is
$R_{n}(x,y;t)=\left(\,I\>-\> K_{n}\,\right)^{-1}\, K_{n}(x,y)$.
 We have a large $n$-expansion of the Hermite kernel $K_{n}(x,y)$,
 therefore to derive an expansion for the resolvent kernel we only
 need to derive an expansion of the kernel $\left(\,I\>-\>
 K_{n}\,\right)^{-1}(x,y)$ and multiply the two operators to have our desired result.
 The first part of this section will be devoted to doing that,
 in the second part we will use that result to derive the expansion
 for $R_{n}$. The third part will derive an
 expansion of $Q_{n}$ and $P_{n}$.

 \subsection{Large $n$-Expansion of $\left(I-
 K_{n}\,\right)^{-1}(x,y)$}
 Let
 \begin{equation}\label{tau}
 \tau(x)\>=\> \sqrt{2(n+c)} \> +\>
 x\,2^{-\frac{1}{2}}\,n^{-\frac{1}{6}}
\end{equation}
be the scaling function and $\chi$ the characteristic function of
the set $(t, \iy)$. \\If $L$ is an integral operator with kernel
$L(x,y)$, we will write $L_{\tau}$ for the scaled integral operator
with kernel
\begin{equation}\label{scaleforkernel}
\frac{1}{\sqrt{2}n^{\frac{1}{6}}}L(\tau(x),\tau(y)).
\end{equation}
With this convention the scaled Hermite kernel \eqref{hermite
kernel} has the following representation.
\begin{eqnarray*}
K_{n}(\tau(X),\tau(y))\,d\tau(x) \>= \>
\tau^{\,_{'}}K_{n}(\tau(X),\tau(y))\>=\>  K_{\airy}(X,Y)
-c\airy(X)\airy(Y) n^{-\frac{1}{3}} +
\end{eqnarray*}
\begin{eqnarray*}
\frac{1}{20}\biggl[(X + Y)\airy^{\prime}(X)\airy ^{\prime}(Y) -
(X^2+XY+Y^2)\airy(X)\airy(Y) +
\end{eqnarray*}
\begin{equation}\label{hermitekernel}
 \left. \frac{-20c^2 +3 }{2}(\airy ^{\prime}(X)\airy(Y) +
\airy(X)\airy ^{\prime}(Y)) \right]n^{-\frac{2}{3}}  +O(n^{-1})
E(X,Y).
\end{equation}
Following Tracy and Widom, we denote the kernel of
$\left(\,I\>-\>K_{n,2}\right)^{-1}$ by $\rho_{n}(x,y)$, the
characteristic function of the set $(t,\iy)$ by $\chi(x)$, and the
scaled function $\chi_{\tau}$ the characteristic function os the set
$(s,\iy)$ as $\tau(s)=t$. Note that
\begin{equation}
\left(\,I\>-\>K_{n,2}\right)^{-1}\>\>=\>\>
\left(\,I\>-\>\chi_{\tau}\>K_{n,2\,\tau}\>\chi_{\tau}\right)^{-1}.
\end{equation}
But $\chi_{\tau}$ scales to the characteristic function of the set
$(s,\iy)$ with $t=\tau(s)$. To simplify notations we will not
mention explicitly $\chi_{(s,\iy)}$, but think of the various
operators as acting on the set $(s,\iy)$. With this in mind, we see
that the kernel of $\left(\,I\>-\>K_{n,2}\right)^{-1}$ is
\begin{equation*}
\rho_{n}(x,y)\>\>=\>\>
\left(\,I\>-\>\frac{1}{2^{\frac{1}{2}}n^{\frac{1}{6}}}K_{n,2}(\,\tau(X)\,
, \, \tau(Y)\,)\right)^{-1}
\end{equation*}
We combine this with  \eqref{hermitekernel} to have
\begin{equation*}
\rho_{n}(x,y) \>=\> \left( \,I\>-\>K_{\airy}(X,Y) +c\airy(X)\airy(Y)
n^{-\frac{1}{3}} - \frac{1}{20}\biggl[(X + Y)\airy^{\prime}(X)\airy
^{\prime}(Y) \right.
\end{equation*}
\begin{equation*}
\left. - (X^2+XY+Y^2)\airy(X)\airy(Y) + \frac{-20c^2 +3 }{2}(\airy
^{\prime}(X)\airy(Y) + \airy(X)\airy ^{\prime}(Y))
\right]n^{-\frac{2}{3}}
\end{equation*}
\begin{equation}
 +O(n^{-1}) E(X,Y) \biggr)^{-1}
\end{equation}
\begin{equation*}
=\> \biggl( \left(\,I\>-\>K_{\airy}(X,Y)\right)\cdot\biggl\{ \,I \>
+ \left(\,I\>-\>K_{\airy}(X,Y)\right)^{-1}\biggl[ \> c
\airy(X)\airy(Y) n^{-\frac{1}{3}} -
\end{equation*}
\begin{equation*}
\frac{1}{20}\biggl[(X + Y)\airy^{\prime}(X)\airy ^{\prime}(Y) -
(X^2+XY+Y^2)\airy(X)\airy(Y) +
\end{equation*}
\begin{equation}\label{eq1}
\left. \left.\left. \left.\frac{-20c^2 +3 }{2}(\airy
^{\prime}(X)\airy(Y) + \airy(X)\airy ^{\prime}(Y))
\right]n^{-\frac{2}{3}}  +O(n^{-1}) E(X,Y)
\right]\right\}\right)^{-1}.
\end{equation}
 We now think of each term in the large bracket as kernel of an integral operator
 on $(s,\iy)$. We know the existence of $\left(\,I\>-\>K_{\airy}\right)^{-1}$. If
we factor out this operator in the last equation, we find that
 \eqref{eq1} can be represented by
\begin{equation*}
 \biggl(  \,I \> +
\left(\,I\>-\>K_{\airy}\right)^{-1}(X,Y)\biggl\{ \> c
\airy(X)\airy(Y) n^{-\frac{1}{3}} - \frac{1}{20}\biggl[(X +
Y)\airy^{\prime}(X)\airy ^{\prime}(Y)
\end{equation*}
\begin{equation*}
 - (X^2+XY+Y^2)\airy(X)\airy(Y) + \frac{-20c^2 +3 }{2}(\airy
^{\prime}(X)\airy(Y) + \airy(X)\airy ^{\prime}(Y))
\biggr]n^{-\frac{2}{3}}
\end{equation*}
\begin{equation}\label{eq2}
   +O(n^{-1}) E(X,Y) \biggr\}\biggr)^{-1}\cdot
\left(\,I\>-\>K_{\airy}\right)^{-1}(X,Y).
\end{equation}
We have the following results (see for example \cite{Trac4}).

If  $M$  denotes multiplication by the independent variable, then
\begin{equation*}\hspace{-1in}\textrm{the integral operator}\>\> M^{i}\airy \otimes M^{j}\airy
\>\> \textrm{has kernel}\>\> X^{i}\airy(X)\,Y^{j}\airy(Y),
\end{equation*}
\begin{equation*}\hspace{-0.8in}\textrm{the integral operator}\>\> M^{i}\airy^{'}\otimes M^{j}\airy
\>\> \textrm{has kernel}\>\>
X^{i}\airy^{'}(X)\,Y^{j}\airy(Y),\>\>\textrm{and}
\end{equation*}
\begin{equation*}\hspace{-1in}\textrm{the integral operator}\>\> M^{i}\airy\otimes M^{j}\airy^{'}
\>\> \textrm{has kernel}\>\> \;X^{i}\airy(X)\,Y^{j}\airy^{'}(Y).
\end{equation*}

If we denote by
\begin{equation*}\rho(X,Y)\quad \textrm{ the kernel of the operator}
\quad (\,I\>-\>K_{\airy})^{-1}
\end{equation*}
on $\;\left(s\>,\> \iy\right),\;$ then using representation
\eqref{Q} we find that the kernel of
\begin{equation}
\left(\,I\>-\>K_{\airy}\right)^{-1}\cdot M^{i}\airy \otimes
M^{j}\airy
\end{equation}
(the dot here represent operator multiplication) is
\begin{equation}
\bigl(\rho(X,Z)\>,\> Z^{i}\airy(Z)\bigr)_{(s\>,\>\iy)}\,
Y^{j}\airy(Y)\>=\> Q_{i}(X;s)\,Y^{j}\airy(Y),
\end{equation}
the kernel of
\begin{equation}
\left(\,I\>-\>K_{\airy}\right)^{-1}\cdot M^{i}\airy^{'} \otimes
M^{j}\airy
\end{equation}
is
\begin{equation}
\bigl(\rho(X,Z)\>,\> Z^{i}\airy^{'}(Z)\bigr)_{(s\>,\>\iy)}\,
Y^{j}\airy(Y)\>=\> P_{i}(X;s)\,Y^{j}\airy(Y),
\end{equation}
and the kernel of
\begin{equation}
\left(\,I\>-\>K_{\airy}\right)^{-1}\cdot M^{i}\airy^{'} \otimes
M^{j}\airy^{'}
\end{equation}
is given by
\begin{equation}
\bigl(\rho(X,Z)\>,\> Z^{i}\airy^{'}(Z)\bigr)_{(s\>,\>\iy)}\,
Y^{j}\airy^{'}(Y)\>=\> P_{i}(X;s)\,Y^{j}\airy^{'}(Y).
\end{equation}
If we substitute these results in \eqref{eq2}, we have
\begin{equation*}
\rho_{n}(x,y) \>=\> \biggl(  \,I \> - \biggl\{ \> - \> c
Q(X;s)\airy(Y) n^{-\frac{1}{3}} +\>
\frac{1}{20}\biggl[P_{1}(X;s)\airy ^{\prime}(Y)+ P(X;s)Y\airy
^{\prime}(Y)
\end{equation*}
\begin{equation*}
 -  Q_{2}(X;s)\airy(Y) - Q_{1}(X;s)Y\airy(Y) -Q(X;s)Y^{2}\airy(Y) + \frac{-20c^2 +3 }{2}
P(X;s)\airy(Y)
\end{equation*}
\begin{equation}\label{eq3}
 +\frac{-20c^2 +3 }{2}Q(X;s)\airy ^{\prime}(Y) \biggr]n^{-\frac{2}{3}}  +O(n^{-1}) E(X,Y) \biggr\}\biggr)^{-1}\cdot
\left(\,I\>-\>K_{\airy}\right)^{-1}(X,Y).
\end{equation}
 We keep the same notation for the error term which is still a
trace class operator since the product of the bounded operator
$(I-K_{\airy})^{-1} $ with the trace class operator $E$ is trace
class.

Note that the operator $n^{-\frac{1}{3}}L$ in the braces in
\eqref{eq3} is a finite sum of finite rank operators, and therefore
a trace class operator. We have the following representation of our
scaled operator
\begin{equation}
\biggl(\,I\>\>-\>\>K_{n,2}\, \biggr)^{-1}\>\>=\>\> \biggl(\,
I\>\>-\>\> n^{-\frac{1}{3}}L\, \biggr)^{-1}\cdot
\biggl(\,I\>\>-\>\>K_{\airy}\,\biggr)^{-1}.
\end{equation}
The trace class limit of the first factor on the right is the
identity operator which is invertible. Then for large $n$ we can
assume that $ (\, I\>-\> n^{-\frac{1}{3}}L\,)$ is also invertible.
This operator therefore admits a convergent (in trace class norm)
Neumann series expansion for large $n$ of the form
\begin{equation}\label{series1}
\biggl(\, I\>\>-\>\> n^{-\frac{1}{3}}L\, \biggr)^{-1} \>=\>
\sum_{k=0}^{\iy} n^{-\frac{k}{3}}L^{k}\>=\>I\>+\>n^{-\frac{1}{3}}L
\>+\>n^{-\frac{2}{3}}L^{2} \>+\> O(n^{-1})E(X,Y).
\end{equation}
We need to find a large $n$-expansion of $n^{-\frac{2}{3}}L^{2}$.
\begin{equation*}
n^{-\frac{2}{3}}L^{2}\>=\>\biggl\{ \> - \> c Q(X;s)\airy(Y)
n^{-\frac{1}{3}} +\> \frac{1}{20}\biggl[P_{1}(X;s)\airy
^{\prime}(Y)+ P(X;s)Y\airy ^{\prime}(Y)
\end{equation*}
\begin{equation*}
 -  Q_{2}(X;s)\airy(Y) - Q_{1}(X;s)Y\airy(Y) -Q(X;s)Y^{2}\airy(Y) + \frac{-20c^2 +3 }{2}
P(X;s)\airy(Y)
\end{equation*}
\begin{equation}\label{eq4}
 +\frac{-20c^2 +3 }{2}Q(X;s)\airy ^{\prime}(Y) \biggr]n^{-\frac{2}{3}}  +O(n^{-1}) E(X,Y)
 \biggr\}^{2}.
\end{equation}
If we use the representation \eqref{un}, we find that this square is
\begin{equation*}
( -c Q(X;s)\airy(Y) n^{-\frac{1}{3}})\cdot( -c Q(X;s)\airy(Y)
n^{-\frac{1}{3}}) \,+ O(n^{-1})E_{1}(X,Y) \
\end{equation*}
\begin{equation*}
=c^{2} Q(X;s)\bigl(\,\airy(Z)\>,\> Q(Z,s)\,
\bigr)_{(s\,,\,\iy)}\,\airy(Y)\, n^{-\frac{2}{3}} +
O(n^{-1})E_{1}(X,Y)
\end{equation*}
\begin{equation}\label{eq33}
=c^{2} Q(X;s)\,u(s)\,\airy(Y)\, n^{-\frac{2}{3}} +
O(n^{-1})E_{1}(X,Y).
\end{equation}
If we substitute \eqref{eq33} in the series expansion
\eqref{series1}, we find that \eqref{eq3} becomes
\begin{equation*}
\tau^{\,_{'}}\rho_{n}(\tau(X),\tau(Y)) \>=\> \biggl(  \,I \>  - \> c
Q(X;s)\airy(Y) n^{-\frac{1}{3}} +\>
\frac{1}{20}\biggl[P_{1}(X;s)\airy ^{\prime}(Y)+ P(X;s)Y\airy
^{\prime}(Y)
\end{equation*}
\begin{equation*}
 -  Q_{2}(X;s)\airy(Y) - Q_{1}(X;s)Y\airy(Y) -Q(X;s)Y^{2}\airy(Y) + \frac{-20c^2 +3 }{2}
P(X;s)\airy(Y)
\end{equation*}
\begin{equation}\label{eq5}
 +\frac{-20c^2 +3 }{2}Q(X;s)\airy ^{\prime}(Y) + 20c^{2}
 Q(X;s)\,u(s)\,\airy(Y)\biggr]n^{-\frac{2}{3}}\;\biggr)\cdot
\left(\,I\>-\>K_{\airy}\right)^{-1}(X,Y).
\end{equation}
\begin{equation*}
+O(n^{-1}) E(X,Y)\cdot \left(\,I\>-\>K_{\airy}\right)^{-1}(X,Y).
\end{equation*}
The notation used for the error term suggests that at each step of
the expansion, we add to the existing error term all the
$O(n^{-1})$-terms and rename the error term by $E(X,Y)$. This is our
first result which we restate as the following Lemma.
\begin{lemma}\label{rho-n}
Let $\rho_{n}$ be the kernel of the operator $(I-K_{n,2})^{-1}$ on
$(t,\>\iy)$,  and $\tau$ the transformation defined by \eqref{tau}.
Then as $n\ra\iy$ with $x=\tau(X)$ and $y=\tau(Y)$,
\begin{equation*}
\rho_{n}(x,y) \>=\> \biggl(  \,I \>  - \> c Q(X;s)\airy(Y)
n^{-\frac{1}{3}} +\> \frac{1}{20}\biggl[P_{1}(X;s)\airy
^{\prime}(Y)+ P(X;s)Y\airy ^{\prime}(Y)
\end{equation*}
\begin{equation*}
 -  Q_{2}(X;s)\airy(Y) - Q_{1}(X;s)Y\airy(Y) -Q(X;s)Y^{2}\airy(Y) + \frac{-20c^2 +3 }{2}
P(X;s)\airy(Y)
\end{equation*}
\begin{equation}\label{eq6}
 +\frac{-20c^2 +3 }{2}Q(X;s)\airy ^{\prime}(Y) + 20c^{2}
 Q(X;s)\,u(s)\,\airy(Y)\biggr]n^{-\frac{2}{3}}\;\biggr)\,
\left(\,I\>-\>K_{\airy}\right)^{-1}(X,Y).
\end{equation}
\begin{equation*}
+O(n^{-1}) E(X,Y)
\end{equation*}
uniformly in $s$.\\ The error term $E$ is the kernel of a trace
class operator on $(s,\>\iy)$. Here $P(X,s)=P_{0}(X,s)$,
$Q(X,s)=Q_{0}(X,s)$, $Q_{1},\> P_{1},$ and $Q_{2}$ are defined in
\eqref{Q} and \eqref{P}
\end{lemma}
\subsection{Large $n$-expansion of $R_{n}(x,y)$}
In this section we will combine Lemma \ref{rho-n} and \eqref{Rn} to
derive an expansion of $R_{n}(x,y)$. \eqref{Rn} says that
\begin{equation}\label{Rn2}
\tau^{\,_{'}}R_{n}(\, \tau(X)\>,\> \tau(Y)\,)\>\>=\>\>
\biggl(\rho_{n}(\tau(X),\tau(Z))\>\>,\>\>
\tau^{\,_{'}}K_{n,2}(\tau(Z),\tau(Y))\biggr)_{(s,\iy)}
\end{equation}
First the action of $(I-K_{\airy})^{-1}$ on \eqref{hermitekernel}
gives
\begin{equation*}
R(X,Y) -cQ(X;s)\airy(Y) n^{-\frac{1}{3}}
+\frac{1}{20}\biggl[P_{1}(X;s)\airy ^{\prime}(Y) +
P(X;s)\,Y\airy^{'}(Y)
\end{equation*}
\begin{eqnarray*}
- Q_{2}(X;s)\airy(Y)-Q_{1}(X;s)Y\airy(Y)-Q(X;s)Y^{2}\airy(Y) +
\end{eqnarray*}
\begin{equation}\label{Rn-0}
 \left. \frac{-20c^2 +3 }{2}( P(X;s)\airy(Y) +
Q(X;s)\airy ^{\prime}(Y)) \right]n^{-\frac{2}{3}}  +O(n^{-1})
E(X,Y).
\end{equation}
Next the action of the first factor in \eqref{eq6} can be computed
as follow:\\
The identity will reproduce \eqref{Rn-0}, the $n^{-\frac{1}{3}}$
term will contribute
\begin{equation}\label{Rn-1}
-cQ(X;s)\bigl(\airy(Z)\,,\,R(Z,Y)\bigr)n^{-\frac{1}{3}}\>\>+\>\>c^{2}Q(X;s)\bigl(\airy(Z)\,
,\, Q(X;s)\bigr)\,\airy(Y)\,n^{-\frac{2}{3}},
\end{equation}
and the $n^{-\frac{2}{3}}$ term will contribute
\begin{equation*}
\frac{1}{20}\biggl[P_{1}(X;s)\,\bigl(\airy
^{\prime}(Z)\,,\,R(X,Y)\bigr)\>+ \>P(X;s)\,\bigl(Z\airy
^{\prime}(Z)\,,\,R(Z,Y)\bigr) -
Q_{2}(X;s)\,\bigl(\airy(Z)\,,\,R(Z,Y)\bigr)
\end{equation*}
\begin{equation*}
 -Q_{1}(X;s)\,\bigl(Z\airy(Z)\,,\,R(Z,Y)\bigr)
  -Q(X;s)\,\bigl(Z^{2}\airy(Z)\,,\,R(Z,Y)\bigr) +
\end{equation*}
\begin{equation*}
 +\frac{-20c^2 +3 }{2}\biggl(P(X;s)\,\bigl(\airy(Z)\,,\,R(Z,Y)\bigr)+Q(X;s)\,\bigl(\airy ^{\prime}(Z)\,,\,R(Z,Y)\bigr)\biggr)
\end{equation*}
\begin{equation}\label{Rn-2}
 + 20c^{2}
 Q(X;s)\,u(s)\,\bigl(\airy(Z)\,,\,R(Z,Y)\bigr)\biggr]n^{-\frac{2}{3}}.
\end{equation}
To evaluate the various inner-product appearing in \eqref{Rn-1} and
\eqref{Rn-2} we will make use of the following representation
$R(X,Y)\>=\>\rho(X,Y)\>-\> \delta(X-Y).$ Thus
\begin{equation*}
\bigl(\airy(Z)\,,\,R(Z,Y)\bigr)\>\>=\>\>-\>\airy(Y) + Q(Y;s),
\end{equation*}
\begin{equation*}
\bigl(\airy ^{\prime}(Z)\,,\,R(Z,Y)\bigr)\>\>=\>\>-\>\airy
^{\prime}(Y) + P(Y;s),
\end{equation*}
\begin{equation*}
\bigl(Z\airy(Z)\,,\,R(Z,Y)\bigr)\>\>=\>\>-\> Y\,\airy(Y) +
Q_{1}(Y;s),
\end{equation*}
\begin{equation*}
\bigl(Z^{2}\airy(Z)\,,\,R(Z,Y)\bigr)\>\>=\>\>-\> Y^{2}\,\airy(Y) +
Q_{2}(Y;s),
\end{equation*}
\begin{equation*}
\bigl(Z\airy ^{\prime}(Z)\,,\,R(Z,Y)\bigr)\>\>=\>\>-\> Y\,\airy
^{\prime}(Y) + P_{1}(Y;s).
\end{equation*}
We substitute these values in \eqref{Rn-1} and \eqref{Rn-2}, then
add all the contributions from \eqref{Rn-0}, \eqref{Rn-1} and
\eqref{Rn-2} to obtain Theorem \ref{resolvent kernel}.\\

\subsection{Large $n$-Expansion of $Q_{n}(x;t)$ and $P_{n}(x;t)$}
In this section we will use Lemma \ref{rho-n} together with the
expansion of $\varphi_{n}(x)$ and $\varphi_{n-1}(x)$
derived\footnote{This is the direct consequence of Theorem 1.1 of
\cite{Choup1}} in \cite{Choup1} to give a large $n$-expansion of
$Q_{n}(x;t)$ and $P_{n}(x;t)$ defined by \eqref{Qn} and \eqref{Pn}
respectively. To obtain $\varphi_{n}(x)$ from Theorem 1.1 of
\cite{Choup1}, we need to make the substitution $c\ra c+\frac{1}{2}$
and the factor in Theorem 1.1 is now $n^{\frac{1}{6}}$. For
$\varphi_{n-1}(x)$, we need to make the substitution $c\ra
c+\frac{3}{2}$ and the factor is now
$n^{\frac{1}{6}}\bigl[1+2^{-1}n^{-\frac{2}{3}}X\bigr]$.
\\
We assume without lost of generalities that $n=2k$ is even, and for
$L_{k}^{\alpha}$ the Laguerre polynomial of degree $k$ and order
$\alpha > -1$  we define
\begin{equation*}
 \varphi_{n}(x)= \biggl(\frac{n}{2}\biggr)^{\frac{1}{4}}\frac{ H_n(x)\,  e^{-x^2/2} }{(2^{n}n! \sqrt{\pi})^{1/2}}
\>\>=\>\> \frac{
k^{\frac{1}{4}}(-1)^{k}2^{2k}(k!)L_{k}^{-\frac{1}{2}}(x^2)\,
e^{-x^2/2} }{(2^{2k}(2k)! \sqrt{\pi})^{1/2}}
\end{equation*}
and
\begin{equation*}
\varphi_{n-1}(x)=\biggl(\frac{n}{2}\biggr)^{\frac{1}{4}}\frac{
H_{n-1}(x)\,  e^{-x^2/2}}{ (2^{n-1}(n-1)! \sqrt{\pi})^{1/2}}
\>\>=\>\>\frac{
k^{\frac{1}{4}}(-1)^{k-1}2^{2k-1}(k-1)!\,x\,L_{k-1}^{\frac{1}{2}}(x^2)\,
e^{-x^2/2} }{(2^{2k-2}(2k-2)! \sqrt{\pi})^{1/2}}.
\end{equation*}
We then have for $x\>=\>\tau(X)$,
\begin{displaymath}
\varphi_{n}(x)=n^{\frac{1}{6}}  \left\{  \airy(X) +
 \frac{(2c-1)}{2} \airy^{\prime}(X) n^{-\frac{1}{3}} +
 \right.
\left[ (10\,c^{2}-10\,c +\frac{3}{2})\, X \airy(X)\right.
 \end{displaymath}
 \begin{equation}\label{phi(n)}
 +\>\> X^2
 \airy^{\prime}(X)\biggr]\frac{n^{-\frac{2}{3}}}{20}
+ O(n^{-1}) \airy(X) \biggr\}
\end{equation}
and
\begin{displaymath}
\varphi_{n-1}(x)=n^{\frac{1}{6}}  \left\{  \airy(X) +
 \frac{(2c+1)}{2} \airy^{\prime}(X) n^{-\frac{1}{3}} +
 \right.
\left[ (10\,c^{2}+10\,c +\frac{3}{2})\, X \airy(X) \right.
 \end{displaymath}
 \begin{equation}\label{phi(n-1)}
+ \>\> X^2
 \airy^{\prime}(X)\biggr]\frac{n^{-\frac{2}{3}}}{20} +  O(n^{-1}) \airy(X) \biggr\}
\end{equation}
Next we apply the scaled operator $(I-K_{n})^{-1}$ to these
functions to have our stated result. First
\begin{displaymath}
(I-K_{\airy})^{-1}\,\varphi_{n}(\tau(X))=n^{\frac{1}{6}}  \left\{
Q(X;s) +
 \frac{(2c-1)}{2} P(X;s) n^{-\frac{1}{3}} +
 \right.
 \end{displaymath}
 \begin{displaymath}
\left[ (10\,c^{2}-10\,c +\frac{3}{2})\, Q_{1}(X;s)\right.
 +\>\> P_{2}(X;s)\biggr]\frac{n^{-\frac{2}{3}}}{20}
+ O(n^{-1}) Q(X;s) \biggr\},
\end{displaymath}
and
\begin{displaymath}
(I-K_{\airy})^{-1}\,\varphi_{n-1}(\tau(X))=n^{\frac{1}{6}}  \left\{
Q(X;s) +
 \frac{(2c+1)}{2} P(X;s) n^{-\frac{1}{3}} +
 \right.
 \end{displaymath}
 \begin{displaymath}
\left[ (10\,c^{2}-10\,c +\frac{3}{2})\, Q_{1}(X;s)\right.
 +\>\> P_{2}(X;s)\biggr]\frac{n^{-\frac{2}{3}}}{20}
+ O(n^{-1}) Q(X;s) \biggr\}.
\end{displaymath}
Preceding in a similar fashion as in the derivation of $R_{n}$, we
make the first factor in the right of the \eqref{eq6} acts on these
last two functions and have Theorem \ref{Q_{n}}. Note that the inner
products here are of the form
\begin{equation*}
\bigl(Z^{i}\airy(Z)\,,\,Q(Z,s)\bigr)\>\>=\>\> u_{i}(s)\quad
\mathrm{and} \quad \bigl(Z^{i}\airy^{'}(Z)\,,\,Q(Z,s)\bigr)\>\>=\>\>
v_{i}(s).
\end{equation*}
To conclude this section, we give the following consequence of
Theorem \ref{Q_{n}}.  If we set $ t\>\>=\>\> \tau(s)$ then as $ n\ra
\iy$

\begin{equation*}
q_{n}(\tau(s))\,=\,Q_{n}(\tau(s);\tau(s))=n^{\frac{1}{6}}\left(
q(s)+ \left[\frac{2c-1}{2}p(s)-c
q(s)u(s)\right]n^{\frac{1}{3}}\right.
\end{equation*}
\begin{equation*}
+\left[(10c^{2}-10c+\frac{3}{2})q_{1}(s)+p_{2}(s) +
(-30c^{2}+10c+\frac{3}{2})q(s) v(s) \right.
\end{equation*}
\begin{equation*}
+  p_{1}(s) v(s) +p(s) v_{1}(s)-q_{2}(s) u(s)-q_{1}(s) u_{1}(s)-q(s)
u_{2}(s)
\end{equation*}
\begin{equation}\label{q_{n}}
+ \left.\left.(-10c^{2}+\frac{3}{2})p(s) u(s) +20c^{2}q(s) u^{2}(s)
\right]\frac{n^{-\frac{2}{3}}}{20} +O(n^{-1})e_{q}(s)\right),
\end{equation}
and
\begin{equation*}
p_{n}(\tau(s))\>=\>P_{n}(\tau(s);\tau(s))=n^{\frac{1}{6}}\left(
q(s)+ \left[\frac{2c+1}{2}p(s)-c
q(s)u(s)\right]n^{\frac{1}{3}}\right.
\end{equation*}
\begin{equation*}
+\left[(10c^{2}+10c+\frac{3}{2})q_{1}(s)+p_{2}(s) +
(-30c^{2}-10c+\frac{3}{2})q(s) v(s) \right.
\end{equation*}
\begin{equation*}
+  p_{1}(s) v(s) +p(s) v_{1}(s)-q_{2}(s) u(s)-q_{1}(s) u_{1}(s)-q(s)
u_{2}(s)
\end{equation*}
\begin{equation}\label{p_{n}}
+ \left.\left.(-10c^{2}+\frac{3}{2})p(s) u(s) +20c^{2}q(s) u^{2}(s)
\right]\frac{n^{-\frac{2}{3}}}{20} +O(n^{-1})e_{p}(s)\right)
\end{equation}
Uniformly in $s$. We use the notation
\begin{equation*}
q_{i}(s)\>\>=\>\>Q_{i}(s;s),\quad p_{i}(s)\>\>=\>\>P_{i}(s;s), \quad
e_{q}(s)\>\>=\>\> E_{Q}(s;s) \quad\textrm{and}\quad
e_{p}(s)\>\>=\>\> E_{P}(s;s)
\end{equation*}
and the subscript $n$ is reserved for functions depending on the
size $n$ of the matrices in consideration.

\section{ Large $n$-Expansion of $F_{n,2}(t)$}
In this section we will use the following Fredholm determinant
representation of the probability distribution function of the
largest eigenvalue $F_{n,2}(t)$ in the GUE$_{n}$ case:
\begin{equation}
F_{n,2}(t)\>\>=\>\> \bP(\lambda_{Max}\>\leq t) \>\>=\>\>
\det(I\>-\>K_{n,2}).
\end{equation}
We also have the following two equations, the proof witch can be
found in \cite{Trac7}
\begin{equation}\label{eq7}
\frac{\partial}{\partial t}\, \log \det(I\>-\>K_{n,2})\>\>=\>\> -\>
R_{n}(t,t;t),
\end{equation}
\begin{equation}\label{eq8}
\frac{\partial}{\partial t}\,R_{n}(t,t;t)\>\>=\>\>
-2\,q_{n}(t)\,p_{n}(t).
\end{equation}
Equation \eqref{eq8} gives
\begin{equation*}
\frac{\partial}{\partial t}\, \log \det(I\>-\>K_{n,2})\>\>=\>\> -\>
2\, \int_{t}^{\iy}q_{n}(x)\,p_{n}(x)\,dx
\end{equation*}
where we used the boundary conditions
$(q_{n}p_{n})(\iy)\>=\>0$. \\
Integration by parts and another use of the boundary conditions
gives
\begin{equation}\label{eq9}
\log \det(I\>-\>K_{n,2})\>\>=\>\> -\> 2\,
\int_{t}^{\iy}\biggl(\int_{y}^{\iy}q_{n}(x)\,p_{n}(x)\,d\,x\>\biggr)d\,y\>\>=\>\>
-\> 2\, \int_{t}^{\iy}\,(x-t)q_{n}(x)\,p_{n}(x)\,d\,x;
\end{equation}
and hence
\begin{thm}\label{F_{n,2}(t)}
\begin{equation}\label{F_{n}}
F_{n,2}(t)\>\>=\>\>\det(I\>-\>K_{n,2})\>\>=\>\> \exp{\biggl(- 2\,
\int_{t}^{\iy}\,(x-t)q_{n}(x)\,p_{n}(x)\,d\,x\biggr)}.
\end{equation}
\end{thm}
Observe that this is the finite $n$ analogue of \eqref{TW}.\\
Now we set $t=\tau(s)$ and $x=\tau(X)$, then
\begin{equation*}
2\, \int_{t}^{\iy}\,(x-t)q_{n}(x)\,p_{n}(x)\,d\,x \>=\>
\int_{s}^{\iy}\,(X-
s)\frac{1}{n^{\frac{1}{3}}}q_{n}(\tau(X))\,p_{n}(\tau(X))\,d\,X
\end{equation*}
With the help of \eqref{q_{n}} and \eqref{p_{n}}, the integrand in
this last equation is now
\begin{equation*}
(x-
s)\frac{1}{n^{\frac{1}{3}}}q_{n}(\tau(x))\,p_{n}(\tau(x))\>\>=\>\>
\end{equation*}
\begin{equation*}
f(x;s)\>\>=\>\>(x-s)\biggl\{\, q^{2}(x)\>+\>
2c\biggl[p(x)q(x)-u(x)q^{2}(x)\biggr]n^{-\frac{1}{3}}\>+\>\biggl[(20c^{2}+3)q(x)q_{1}(x)+
\end{equation*}
\begin{equation*}
2p_{2}(x)q(x) +(-60c^2 +3)q^{2}(x)v(x) +2p_{1}(x)q(x)v(x)
+2p(x)q(x)v_{1}(x)-2q_{2}(x)q(x)u(x)
\end{equation*}
\begin{equation*}
-2q_{1}(x)q(x)u_{1}(x)-2q^{2}(x)u_{2}(x) +(-60c^2 +3)p(x)q(x)u(x)
+60c^{2}q^{2}(x)u^{2}(x)
\end{equation*}
\begin{equation}\label{integrand}
+(20c^{2}-5)p^{2}(x) \biggr]\frac{n^{-\frac{2}{3}}}{20}
+O(n^{-1})e(x) \biggr\},
\end{equation}
or
\begin{equation*}
f(x;s)\>\>=\>\>(x-s)\biggl(\, q^{2}(x)\>+\> a(x)n^{-\frac{1}{3}}\>
+\> b(x)\frac{n^{-\frac{2}{3}}}{20} \>+\>O(n^{-1})e(x) \biggr).
\end{equation*}
where
\begin{equation}
a(x)\>=\> 2c\bigl[p(x)q(x)-u(x)q^{2}(x)\bigr]
\end{equation}
and
\begin{equation*}
b(x)\>=\>\biggl[(20c^{2}+3)q(x)q_{1}(x)+2p_{2}(x)q(x) +(-60c^2
+3)q^{2}(x)v(x) +2p_{1}(x)q(x)v(x)
\end{equation*}
\begin{equation*}
+2p(x)q(x)v_{1}(x)-2q_{2}(x)q(x)u(x)-2q_{1}(x)q(x)u_{1}(x)-2q^{2}(x)u_{2}(x)
+60c^{2}q^{2}(x)u^{2}(x)
\end{equation*}
\begin{equation}\label{b(x)}
+(-60c^2 +3)p(x)q(x)u(x)+(20c^{2}-5)p^{2}(x) \biggr].
\end{equation}
We use $x$ instead of $X$  here to simplify notation since $X$ is
just a variable of integration. We therefore have
\begin{equation*}
\det(I\>-\>K_{n,2})= \exp{\biggl(- \hspace{-0.07in}\int_{s}^{\iy}
\hspace{-0.15in} (x-s)\biggl( q^{2}(x)+ a(x)n^{-\frac{1}{3}} +
b(x)\frac{n^{-\frac{2}{3}}}{20} +O(n^{-1})e(x) \biggr)d\,x\biggr)} =
\end{equation*}
\begin{equation*}
\exp{\biggl(- \hspace{-0.07in}\int_{s}^{\iy}\hspace{-0.15in}(x-s)
q^{2}(x)dx\biggr)}\exp{\biggl(-
\hspace{-0.07in}\int_{s}^{\iy}\hspace{-0.15in}(x-s) a(x)dx\,
n^{-\frac{1}{3}}\biggr)}\exp{\biggl(- \hspace{-0.07in}
\int_{s}^{\iy}\hspace{-0.15in}(x-s) b(x)dx\,
\frac{n^{-\frac{2}{3}}}{20}\biggr)} \cdot
\end{equation*}
\begin{equation*}
\exp{\biggl(- \hspace{-0.07in} \int_{s}^{\iy}\hspace{-0.15in}(x-s)
e(x)dx\, O(n^{-1})\biggr)}\>\>=
\end{equation*}
\begin{equation*}
\biggl(1- \int_{s}^{\iy}\hspace{-0.15in}(x-s) a(x)dx\,
n^{-\frac{1}{3}} \>+\> \frac{1}{2}
\biggl[\int_{s}^{\iy}\hspace{-0.15in}(x-s) a(x)dx
\biggr]^{2}n^{-\frac{2}{3}}\>+\> E_{a}(s)\,O(n^{-1})\,\biggr)\cdot
\end{equation*}
\begin{equation*}
\biggl(1 - \int_{s}^{\iy}\hspace{-0.15in}(x-s) b(x)dx\,
\frac{n^{-\frac{2}{3}}}{20}\>+\> E_{b}(s) \,O(n^{-1})\biggr) \cdot
\end{equation*}
\begin{equation*}
\biggl(1- E_{e}(x)\, O(n^{-1})\biggr)\exp{\biggl(-
\int_{s}^{\iy}\hspace{-0.15in}(x-s) q^{2}(x)dx\biggr)}
\end{equation*}
\begin{equation*}
= \> \> \biggl\{1- \int_{s}^{\iy}\hspace{-0.15in}(x-s) a(x)dx\,
n^{-\frac{1}{3}} \>+\>\biggl( 10
\biggl[\int_{s}^{\iy}\hspace{-0.15in}(x-s) a(x)dx \biggr]^{2}-
\int_{s}^{\iy}\hspace{-0.15in}(x-s) b(x)dx
\biggr)\frac{n^{-\frac{2}{3}}}{20}
\end{equation*}
\begin{equation*}
+E_{F}(s)O(n^{-1})\biggr\}\exp{\biggl(-
\int_{s}^{\iy}\hspace{-0.15in}(x-s) q^{2}(x)dx\biggr)},
\end{equation*}
where $E_{a}(s),\>\> E_{b}(s), \>\> \textrm{and}\>\> E_{e}(s)$ are
the reminder when expanding the exponential functions and $E_{F}(s)$
is the collection of all the $O(n^{-1})$ terms. The second factor of
this last equality is known as the Tracy-Widom distribution. The
first factor will be the focus on the reminder of this paper. First
we will find a simplification for the $n^{-\frac{1}{3}}$ term, then
a simplification of the $n^{-\frac{2}{3}}$ term. The error term is a
consequence of our asymptotic.
\subsection{The $n^{-\frac{1}{3}}$ term}
We will use the displayed formulas on page 6 of \cite{Trac4} to
simplify this factor. First note that if we integrate by parts and
use the boundary conditions on $a(x)$,
\begin{equation}\label{integration by parts}
\int_{s}^{\iy}\hspace{-0.15in}(x-s) a(x)dx
\>\>=\>\>\int_{s}^{\iy}\biggl(\int_{y}^{\iy}\hspace{-0.15in}a(x)\,dx\biggr)
dy\,.
\end{equation}
We have
\begin{equation*}
- \int_{s}^{\iy}\biggl(\int_{y}^{\iy}\hspace{-0.15in}a(x)\,dx\biggr)
dx
\>\>=\>\>-c\int_{s}^{\iy}\biggl(\int_{y}^{\iy}\hspace{-0.15in}2q(x)(p(x)-u(x)q(x))\,dx\biggr)
dy
\end{equation*}
\begin{equation*}
=\>\>-c\int_{s}^{\iy}\biggl(\int_{y}^{\iy}\hspace{-0.15in}2q(x)q^{_{'}}(x)\,dx\biggr)
dy
\>\>=\>\>-c\int_{s}^{\iy}\biggl(\int_{y}^{\iy}\hspace{-0.15in}(q^{2}(x))^{'}\,dx\biggr)
dy
\end{equation*}
\begin{equation}\label{first term}
=\>\> -c\int_{s}^{\iy}(-q^{2}(y))\, dy \>\>=\>\>
-c\int_{s}^{\iy}u^{_{'}}(y)\, dy \>\>=\>\> c\,u(s)\,.
\end{equation}
Note that this is the $n^{-\frac{1}{3}}$ term from \cite{Choup1}.

\subsection{The $n^{-\frac{2}{3}}$ term}
We will simplify $b(x)$ in two steps. In the first part we will
simplify the expression containing the constant $c$, and in the
second step simplify the remaining expression.

The expression proportional to the constant $c$ is
\begin{equation}\label{c term}
 10 \biggl[\int_{s}^{\iy}\hspace{-0.15in}(x-s) a(x)dx
\biggr]^{2}- 20c^{2}\int_{s}^{\iy}\hspace{-0.15in}(x-s)
\bigl(qq_{1}-3q^{2}v+3q^{2}u^{2}-3pqu+p^{2}\bigr)(x)dx
\end{equation}
Equation \eqref{first term} says that the first term is
$10c^{2}u^{2}(s)$. Equation (2.12) of \cite{Trac7} together with our
our definition of $q_{i}$ give $q_{1}(s)\>=\>
sq(s)-v(s)q(s)+u(s)p(s)$. If we substitute this expression of
$q_{1}(s)$ in the second term of \eqref{c term}, then it becomes
\begin{equation*}
\int_{s}^{\iy}\hspace{-0.15in}(x-s)
\bigl(xq^{2}(x)-4q^{2}(x)v(x)+3q^{2}(x)u^{2}(x)-2p(x)q(x)u(x)+p^{2}(x)\bigr)dx
\end{equation*}
\begin{equation*}
=\>\int_{s}^{\iy}\int_{y}^{\iy}\hspace{-0.15in}
\bigl(xq^{2}(x)-4q^{2}(x)v(x)+3q^{2}(x)u^{2}(x)-2p(x)q(x)u(x)+p^{2}(x)\bigr)dx\,dy.
\end{equation*}
In the following steps we integrate this last expression
\begin{equation*}
xq^{2}(x)-4q^{2}(x)v(x)+3q^{2}(x)u^{2}(x)-2p(x)q(x)u(x)+p^{2}(x)
\end{equation*}
\begin{equation*}
=\>\>-2u(x)p(x)q(x)+2u^{2}(x)q^{2}(x)+u^{2}(x)q^{2}(x)-2v(x)q^{2}(x)-2v(x)q^{2}(x)+xq^{2}(x)
\end{equation*}
\begin{equation*}
-2v(x)q^{2}(x)+xq^{2}(x)+u(x)p(x)q(x)-u(x)p(x)q(x)+p^{2}(x)
\end{equation*}
\begin{equation*}
=
\>\>-2\bigl(p(x)-q(x)u(x)\bigr)q(x)u(x)+q^{2}(x)\bigl(u^{2}(x)-2v(x)\bigr)+q(x)\bigl(xq(x)-2v(x)q(x)+p(x)u(x)\bigr)
\end{equation*}
\begin{equation*}
+q(x)\bigl(xq(x)-2v(x)q(x)+p(x)u(x)\bigr)+p(x)\bigl(p(x)-q(x)u(x)\bigr)
\end{equation*}
\begin{equation*}
=\>\>-2q^{'}(x)q(x)u(x)+q^{2}(x)q^{2}(x)+q(x)p^{'}(x)+p(x)q^{'}(x)
\end{equation*}
\begin{equation*}
=\>\>\bigl(-q^{2}(x)\bigr)^{'}u(x)-q^{2}(x)u^{'}(x)+\bigl(p(x)q(x)\bigr)^{'}\>=
\>\bigl(-q^{2}(x)u(x)\bigr)^{'}+\bigl(p(x)q(x)\bigl)^{'}
\end{equation*}
\begin{equation}\label{check for c term}
=\>\> \bigl(\frac{1}{2}u^{2}(x)\bigl)^{''}-v^{''}(x).
\end{equation}
The second integral in  \eqref{c term} is therefore,
\begin{equation}
20c^{2}v(s)-10c^{2}u^{2}(s).
\end{equation}
This last expression is due to the fact that the functions $ u_{i},
\> u_{i}^{'},\> v_{i}$ and $v_{i}^{'}$ are zero at infinity. The
derivation of the various integrals used for \eqref{check for c
term} can be found in \cite{Trac4}. We showed that the term
containing the constant $c$ simplifies to
\begin{equation}\label{final c term}
20c^{2}v(s).
\end{equation}
Note that this is the same term derived in \cite{Choup1}.

In similar way we show that
\begin{equation*}
\int_{s}^{\iy}\int_{y}^{\iy}\bigl(-3q_{1}q+3q^{2}v-3p^{2}+3upq\bigr)(x)\,dx\,dy
\end{equation*}
\begin{equation*}
=\>\>\int_{s}^{\iy}\int_{y}^{\iy}\bigl(-3xq^{2}+6q^{2}v-3upq
+3upq-3p^{2}\bigr)(x)\,dx\,dy
\end{equation*}
\begin{equation*}
=\>\int_{s}^{\iy}\int_{y}^{\iy}\bigl(-3q[xq-2q^{2}v+up]
-3p[-uq+p]\bigr)(x)\,dx\,dy
\end{equation*}
\begin{equation}\label{v term}
=\>-3\int_{s}^{\iy}\int_{y}^{\iy}\bigl(qp^{'}+pq^{'}
\bigr)(x)\,dx\,dy=3\int_{s}^{\iy}\int_{y}^{\iy}\bigl(v(x)\bigr)^{''}\,dx\,dy\>=\>3v(s).
\end{equation}
Suppose that $L$ is the integral of $l$ subject to the boundary
condition $L(\iy)=0$, then
\begin{equation}\label{by parts}
-\int_{s}^{\iy}(x-s)l(x)\,dx = \int_{s}^{\iy}L(x)\,dx.
\end{equation}
Using the following representation (the derivation of which can be
found in \cite{Trac7},)
\begin{equation*}
\bigl(-q_{1}q+q^{2}v-p^{2}+upq\bigr)(x)\>=\>
\bigl(u_{1}-uv+w\bigr)^{'}(x),
\end{equation*}
we find that equation \eqref{v term} reduces to the following
integral
\begin{equation}
\int_{s}^{\iy}\bigl(u_{1}-uv+w\bigr)(x) \,dx = -v(s),
\end{equation}
or that
\begin{equation}\label{u1}
u_{1}(s)-u(s)v(s)+w(s)  = v^{'}(s)=-p(s)q(s).
\end{equation}

At this stage of the simplification the $n^{-\frac{2}{3}}$ term is
\begin{equation*}
(20c^{2}-3)\,v(s)
-\int_{s}^{\iy}(x-s)\bigl(6qq_{1}+2p_{2}q+2p_{1}qv+2pqv_{1}-2q_{2}qu-2q_{1}qu_{1}-2q^{2}u_{2}-2p^{2}\bigr)(x)\,dx
\end{equation*}
\begin{equation*}
=\>\> (20c^{2}-3)\,v(s)\>\> -\int_{s}^{\iy}(x-s)h(x)\,dx.
\end{equation*}
If we note that
\begin{equation*}
h(x)=
\bigl(-6u_{1}^{'}-2v_{2}^{'}-2v_{1}^{'}v-2v^{'}v_{1}+2u_{2}^{'}u+
\bigl(u_{1}^{2}\bigr)^{'}+2u^{'}u_{2}+2w^{'}\bigr)(x)
\end{equation*}
\begin{equation*}
= \>\>\bigl(-6u_{1}-2v_{2}-2v_{1}v+2u_{2}u+
u_{1}^{2}+2w\bigr)^{'}(x),
\end{equation*}
then the $n^{-\frac{2}{3}}$ term becomes
\begin{equation}
\>\> (20c^{2}-3)\,v(s)\>\>
+\int_{s}^{\iy}\bigl(-6u_{1}-2v_{2}-2v_{1}v+2u_{2}u+
u_{1}^{2}+2w\bigr)(x)\,dx.
\end{equation}
This is where we stop our simplification of this term. To fully
simplify this result to match equation \eqref{GUE Edgeworth}, we
need to derive\footnote{The derivation  was simple for the term
containing the constant $c$ since we can trace it out.} new integral
similar to equation \eqref{u1} from the following representation,
\begin{equation}\label{matching}
\int_{s}^{\iy}\bigl(6u_{1}+2v_{2}+2v_{1}v-2u_{2}u-
u_{1}^{2}-2w\bigr)(x)\,dx \>=\>2w_{1}-3u_{2}  +
u_{1}v_{0}-u_{0}\tilde{v}_{1}
\end{equation}

 We
find that the large $n$-expansion of the probability distribution
function of the largest eigenvalue for the GUE$_{n}$ case is given
by the following; if we set
\begin{equation}
\tau(s)=\sqrt{2(n+c)} +\frac{s}{2^{\frac{1}{2}}n^{\frac{1}{6}}},
\end{equation}
then as $n\rightarrow \iy$,  we have
\begin{equation*}
\mathrm{F}_{n,2}\bigl(\tau(s)\bigr)= \mathrm{F}_{2}(s) \biggl\{
1\>\>+\>\>c\,u(s)\,n^{-\frac{1}{3}}\>\> +
\end{equation*}
\begin{equation}\label{f_{n,2}}
\frac{n^{-\frac{1}{3}}}{20}\biggl[(20c^{2}-3)\,v(s)\>\>
+\int_{s}^{\iy}\bigl(-6u_{1}-2v_{2}-2v_{1}v+2u_{2}u+
u_{1}^{2}+2w\bigr)(x)\,dx\biggr] \biggr\} +O(n^{-1})
\end{equation}
uniformly is $s$.
\section{Conclusion}
Our motivation in this paper was to find large $n$-expansion of
$q_{n}$ and $p_{n}$. The importance of such large $n$ expansions is
that they not only give a direct proof of Theorem \ref{GUE}
(GUE$_{n}$ case), but they are essential ingredients in extending
Theorem \ref{GUE} to the GOE$_{n}$ and GSE$_{n}$ cases. We will
return to these cases in a subsequent paper.

\clearpage \vspace{3ex} \noindent\textbf{\large Acknowledgements: }
The author would like to thank Professor Craig Tracy for the
discussions that initiated this work and for the invaluable
guidance, and the Department of Mathematical Sciences at the
University of Alabama in Huntsville.

%

\end{document}